\documentclass[showpacs,preprintnumbers,amsmath,amssymb,a4paper,superscriptaddress]{revtex4}
\usepackage{amsmath}
\usepackage{graphicx}
\usepackage{dcolumn}
\usepackage{bm}
\usepackage{booktabs}
\usepackage{calc}
\usepackage{multirow}
\usepackage{mathtools}

\def\nn{\nonumber}

\DeclareMathAlphabet{\mathpzc}{OT1}{pzc}{m}{it}

\newcommand{\vek}[1]{\mathbf{#1}}
\newcommand {\beq} {\begin{eqnarray}}
\newcommand {\eeqn} [1] {\label{#1} \end{eqnarray}}

\newcommand{\rel}[1]{\rho\,}
\newcommand{\ra}[1]{\hat{\mathbf{r}}\,}
\newcommand{\ri}[1]{\mathbf{R}\,}
\newcommand{\ma}[1]{m\,}
\newcommand{\mi}[1]{M\,}
\newcommand{\xa}[1]{x}
\newcommand{\ya}[1]{y\,}
\newcommand{\za}[1]{z\,}
\newcommand{\X}[1]{X\,}
\newcommand{\Y}[1]{Y\,}
\newcommand{\Z}[1]{Z\,}

\begin{document}

\title{Acceleration and twisting of neutral atoms by strong elliptically polarized short-wavelength laser pulses}
\date{\today}
\pacs{32.60.+i,33.55.Be,32.10.Dk,33.80.Ps}
\author{Vladimir S.Melezhik}
\email[]{melezhik@theor.jinr.ru}
\affiliation{Bogoliubov Laboratory of Theoretical Physics, Joint Institute for Nuclear Research, 6 Joliot-Curie St., Dubna, Moscow Region 141980, Russian Federation}
\affiliation{Dubna State University, 19 Universitetskaya St., Dubna, Moscow Region 141982, Russian Federation}
\author{Sara Shadmehri}
\email[]{shadmehri@jinr.ru}
\affiliation{Meshcheryakov Laboratory of Information Technologies, Joint Institute for Nuclear Research, 6 Joliot-Curie St., Dubna, Moscow Region 141980, Russian Federation}
\date{\today}

\begin{abstract}\label{txt:abstract}
We have investigated non-dipole effects in the interaction of a hydrogen atom with elliptically polarized laser pulses of intensity 10$^{14}$ W/cm$^2$ with about 8 fs duration. The study was performed within the framework of a hybrid quantum-quasiclassical approach in which the time-dependent Schr\"odinger equation for an electron and the classical Hamilton equations for the center-of-mass (CM) of an atom are simultaneously integrated. It is shown that the spatial inhomogeneity $ \mathbf{k}\cdot\mathbf{r}$  of the laser field and the presence of a magnetic component in it lead to the non-separability of the CM and electron variables in a neutral atom and, as a consequence, to its acceleration. We have established a strict correlation between the total probability of excitation and ionization of an atom and the velocity of its CM acquired as a result of interaction with a laser pulse. The acceleration of the atom weakly depends on the polarization of the laser in the considered region (\mbox{5 eV $\lesssim \hbar\omega \lesssim $ 27 eV})  of its frequencies. However, the transition from linear to elliptical laser polarization leads to the twisting of the atom relative to the axis directed along the pulse propagation (coinciding with the direction of the momentum of the accelerated atom). It is shown that with increasing ellipticity the twisting effect increases and reaches its maximum value with circular polarization. At this point  the projection of the orbital angular momentum acquired by the electron onto the direction of the pulse propagation reaches its maximum value. Further exploration of the possibilities for producing accelerated and twisted atoms with electromagnetic pulses is of interest for a number of prospective applications.

\end{abstract}

\maketitle

\section{INTRODUCTION}

Non-dipole corrections in the interaction of an atom with laser radiation that arise when taking into account the spatial inhomogeneity $\mathbf{k}\cdot\mathbf{r}$ of the electromagnetic wave and the presence of a magnetic component in it lead to ``entanglement'' of the variables of the center-of-mass (CM) and electrons in a neutral atom and, as a consequence, to its acceleration \cite{Mel2023}. We studied this effect in \cite{MS2023}, as well as the accompanying processes of excitation and ionization of the hydrogen atom in strong (10$^{12}$ - 2$\times$10$^{14}$) W/cm$^2$ linearly polarized short-wave (5 eV $\leq\hbar\omega \leq$ 27 eV) short-range electromagnetic pulses with a duration of about 8 fs. The study was carried out within the framework of a hybrid quantum-quasiclassical approach (see \cite{Mel2023} and references therein), in which the time-dependent Schr\"odinger equation for the electron and the classical Hamilton equations for the CM of the atom with a close coupling between them are simultaneously integrated. A strong correlation was discovered between the velocity (momentum) of the CM of the atom at the end of the laser pulse and the total probability $P_{ex}+P_{ion}$ of excitation and ionization of the atom. Two mechanisms of atomic acceleration were established in consequence of single-photon and two-photon excitation of the atom.  It was shown that the one-photon mechanism leads to a linear dependence of the atomic velocity at the end of the laser pulse on the laser intensity, and the two-photon mechanism leads to a quadratic dependence. Optimal conditions for the frequency and intensity of the electromagnetic wave were found for the acceleration of atoms without their noticeable ionization in the studied range of changes in laser parameters \cite{MS2023}.

In this paper we extend our consideration to the case of elliptical polarization of the laser pulse. Within the framework of the developed approach\cite{Mel2023,MS2023}, the influence of laser ellipticity on the acceleration of a hydrogen atom and the accompanying effects of its excitation and ionization are studied. We have found that a strict correlation between the total probability of excitation and ionization of an atom and the velocity of its CM acquired as a result of interaction with a laser pulse is preserved for any ellipticity. In this case, the acceleration of the atom depends weakly on the polarization of the laser in the considered region (\mbox{5 eV $\lesssim \hbar\omega \lesssim $ 27 eV})  of its frequencies. However, it is established that the deviation from the linear polarization of the electromagnetic pulse leads to the twisting of the atom. Moreover, the effect increases with increasing ellipticity and reaches a maximum with circular polarization. At this point the projection of the orbital angular momentum acquired by the electron onto the direction of pulse propagation reaches its maximum value. In this regard, it should be noted that the physics of twisted photons~\cite{photon} and electrons~\cite{electron} is currently one of the hot research areas due to its potentially interesting applications here (see \cite{Serbo,Bliokh} and references therein). For example, electron vortex beams were used to study chirality, magnetization mapping and transfer of angular momentum to nanoparticles \cite{Bliokh}. Several proposals were made to create vortex beams of composite particles (neutrons, protons, and atoms)~\cite{Clark}. It is supposed that the twisting of composite particles can be used to alter the fundamental interactions of such particles and enable probing their internal structure. However, until recently only one successful experiment of creating a vortex beam of atoms has been realized: in the \cite{Luski}, a beam of twisted helium atoms was obtained with a fork diffraction grating. Here we discusses the possibility of producing twisted accelerated atoms by using elliptically polarized laser pulses.

In the next section, our theoretical approach and the principal elements of the computational scheme are given. In Section III, the results and discussions are presented. The concluding remarks are given in the last section. Some technical details of the computations are discussed in the appendices.

\section{PROBLEM FORMULATION AND COMPUTATIONAL SCHEME}

We consider the dynamics of a hydrogen atom interacting with an elliptically polarized laser pulse determined by the vector potential (atomic units $e^2=m_e=\hbar=1$ are used hereafter except where otherwise noted)
\begin{eqnarray}
{\bf A}=-\frac{E_0f(t)}{\omega\sqrt{1+\varepsilon^2}}\Big[{\bf\hat{x}}\sin(\omega t-\mathbf{k}\cdot\mathbf{r})-\varepsilon{\bf\hat{y}}\cos(\omega t-\mathbf{k}\cdot\mathbf{r})\Big]\,,
\end{eqnarray}
where the pulse envelope
\begin{eqnarray}
f(t)=\sin^2(\frac{\pi t}{NT})~~,~~~0\leq t \leq T_{out}=NT=100\pi \nonumber
\end{eqnarray}
contains $N$ optical cycles of the time period $T=2\pi/\omega$ defined by the laser frequency $\omega$. Here $E_0$ is the strength of the field defined by the laser intensity $I=\epsilon_0cE^2_0/2$ ($\epsilon_0$ is the vacuum permittivity), ${\bf k}=k{\bf \hat{z}}=\omega/c{\bf \hat{z}}$ and $c=137$ are the wave vector and the speed of light, respectively. This pulse propagates in the $z$-direction and is polarized in the $xy$-plane.
The limiting cases $\varepsilon=0$ and $\varepsilon=1$ of the ellipticity $0 \leq \varepsilon \leq 1$ correspond to the linearly and circularly polarized fields, respectively. In our investigation, the pulse duration was fixed as in our previous work~\cite{SM2023} by $T_{out}=NT=100\pi$a.u.$\approx$7.6fs, which is defined for varying laser frequency $\omega$ by varying the number of optical cycles $N$.

The electric ${\bf E}=-\frac{d{\bf A}}{dt}$ and magnetic ${\bf B}={\bf \nabla}\times{\bf A}$ fields of the laser pulse take the forms

\begin{eqnarray}
{\bf E}=E_0(t)\Bigg\{{\bf\hat{x}}\Big[\cos(\omega ( t-\frac{z}{c}))+\frac{\bar{f}(t)}{2Nf(t)}\sin(\omega (t-\frac{z}{c}))\Big]
+\varepsilon {\bf\hat{y}}\Big[\sin(\omega (t-\frac{z}{c}))-\frac{\bar{f}(t)}{2Nf(t)}\cos(\omega (t-\frac{z}{c}))\Big]\Bigg\}
\end{eqnarray}
\begin{eqnarray}
{\bf B}=\frac{1}{c}E_0(t)\left[{\bf\hat{y}}\cos(\omega (t-\frac{z}{c}))-\varepsilon{\bf\hat{x}}\sin(\omega (t-\frac{z}{c}))\right]~,
\end{eqnarray}
where $\bar{f}(t)=\sin(\frac{2\pi t}{NT})$ and the common factor $E_0(t)=E_0f(t)/\sqrt{1+\varepsilon^2}$ in the formulas above is defined by the field strength $E_0$, the form of the field envelope $f(t)$, and the ellipticity $\varepsilon$.

Usually, the interaction of the hydrogen atom with laser fields is considered in the dipole approximation
\begin{eqnarray}
V({\bf r},t)= E_0(t)\Bigg\{\Big[\cos(\omega t)+\frac{\bar{f}(t)}{2Nf(t)}\sin(\omega t)\Big]x
+\varepsilon \Big[\sin(\omega t)-\frac{\bar{f}(t)}{2Nf(t)}\cos(\omega t)\Big]y\Bigg\}\,,
\end{eqnarray}
in which the magnetic component (3) ($\sim 1/c=1/137$) and the spatial dependence in the propagation direction of the pulse ($\sim kz=\omega z/c= \omega z/137$) in (2)  are neglected. Here $x$ and $y$ are the components of the relative variable ${\bf r}= x{\bf \hat{x}}+y{\bf \hat{y}}+z{\bf \hat{z}}={\bf r}_e-{\bf r}_p$ of the electron and proton in the hydrogen atom, where ${\bf r_e}$ and ${\bf r_p}$ are the electron and proton variables, respectively.

Going beyond the dipole approximation, i.e. accounting for the spatial inhomogeneity $\mathbf{k}\cdot\mathbf{r}$ of the vector potential (1),  leads to the following modification of the interaction potential (see Appendix A):
\begin{align}
\label{eq:v0}
V({\bf r},t)\Rightarrow V({\bf r},t) + V_1({\bf r},t)+V_2({\bf r},{\bf R},t)~,
\end{align}
where
\begin{eqnarray}
\label{eq:v1}
V_1({\bf r},t) =
\frac{E_0(t)}{c}\Bigg\{\omega\Big[\sin(\omega t)-\frac{\bar{f}(t)}{2Nf(t)}\cos(\omega t)\Big] zx
-\varepsilon\omega\Big[\cos(\omega t)+\frac{\bar{f}(t)}{2Nf(t)}\sin(\omega t)\Big] zy
 + \Big[\cos(\omega t)\hat{l}_y - \varepsilon \sin(\omega t)\hat{l}_x \Big]\Bigg\} \,\,,
\end{eqnarray}
and
\begin{align}
\label{eq:v2}
V_2({\bf r},{\bf R},t) = \frac{E_0(t)}{c}
\Bigg\{\omega \Big[ \sin(\omega t)-\frac{\bar{f}(t)}{2Nf(t)}\cos(\omega t)\Big](zX +xZ)
-\varepsilon \omega \Big[\cos(\omega t)+\frac{\bar{f}(t)}{2Nf(t)}\sin(\omega t)\Big](zY +yZ) \nonumber\\
~+\Big[\cos(\omega t)(Z\hat{p}_x-X\hat{p}_z)+\varepsilon \sin(\omega t)(Z\hat{p}_y-Y\hat{p}_z)\Big]\Bigg\} \,.
\end{align}
This potential is written in the center-of-mass ${\bf R}=X{\bf\hat{x}}+Y{\bf\hat{y}}+Z{\bf\hat{z}}$ and relative ${\bf r}$ variables, where $\hat{l}_x$ and $\hat{l}_y$ in (\ref{eq:v1}) are the $x$ and $y$ components of the electron orbital angular momentum relative proton. In deriving these formulas, we have neglected the terms $\sim 1/c^2$ and $\sim 1/M=1/(m_e+m_p)$ and higher orders (see Appendix A).
Thus, the total Hamiltonian of the hydrogen atom in the laser field takes the form
\begin{align}
H({\bf r},{\bf R},t) = \frac{{\bf \hat{P}}^2}{2M} + h_0({\bf r}) +V({\bf r},t)+ V_1({\bf r},t) +V_2({\bf r},{\bf R},t)~,
\end{align}
where the $h_0$-Hamiltonian
\begin{align}
h_0({\bf r})=\frac{{\bf \hat{p}}^2}{2\mu} -\frac{1}{r}
\end{align}
describes the relative motion of an electron and a proton in the Coulomb field between them. Here ${\bf \hat{p}}$ is the momentum operator of the relative motion of the electron with respect to the proton, ${\bf \hat{P}}$ is the momentum of the CM, $\mu=m_em_p/(m_e+m_p)$ is the reduced mass of the atom, and $M=m_e+m_p$. The term $V({\bf r},t)$ describes the interaction of the atom with the laser pulse in the dipole approximation, and two additional terms $V_1({\bf r},t)$ and $V_2({\bf r},{\bf R},t)$ describe the corrections to the dipole approximation of the order of $\sim 1/c$ and $\sim \omega/c$. Note that the last term $V_2({\bf r},{\bf R},t)$ in the total Hamiltonian entangles the CM and electron variables and leads to the non-separability of the problem. It should also be noted that although the importance of non-dipole corrections in strong laser fields was recognized quite a long time ago~\cite{Reiss,Kylstra,Hammers,Forre}, the non-separability of the CM in strong fields has been little studied to date due to the complexity of this problem~\cite{Bray,Mel2023}.

Following the computational scheme suggested and developed in \cite{Mel2023,MS2023}, we quantitatively investigate the problem of the hydrogen atom in the elliptically polarized laser field described by the Hamiltonian (8) within the quantum-quasiclassical approach in which the quantum dynamics of the electron relative the proton is described by the time-dependent 3D Schr\"odinger equation
\begin{align}
\label{eq:Schr}
i\frac{\partial}{\partial t}\psi(\vek{r},t) = \Big[h_0(\vek{r}) + V(\vek{r},t)
+ V_1(\vek{r},t)+V_2(\vek{r},\vek{R}(t),t)\Big]\psi(\vek{r},t) \,,
\end{align}
which is integrated simultaneously with the classical Hamiltonian equations
\begin{align}
\label{eq:Hamilton}
\frac{d}{d t}\vek{P}& = -\frac{\partial}{\partial \vek{R}}H_{eff} (\vek{P},\vek{R},t)\nonumber\\
\frac{d}{d t}\vek{R}& = \frac{\partial}{\partial \vek{P}}H_{eff} (\vek{P},\vek{R},t)
\end{align}
describing the CM dynamics and its close coupling with the 3D Schr\"odinger equation (\ref{eq:Schr}) via the mixing term $V_2(\vek{r},\vek{R}(t),t)$ in the Hamiltonian of Eq.(\ref{eq:Schr}) (which depends on $\vek{R}(t)$ parametrically) and in the effective  Hamiltonian
\begin{eqnarray}
\label{eq:eff}
H_{eff}\left({\mathbf P},{\mathbf R}\right)=\frac{{\mathbf P}^2}{2M}+\langle \psi({\mathbf r},t)\vert V_2({\mathbf r},{\mathbf R}(t),t)\vert \psi({\mathbf r},t) \rangle \,\,,
\end{eqnarray}
in the classical equations (\ref{eq:Hamilton}).

The application of the quantum-quasiclassical computational scheme (\ref{eq:Schr},\ref{eq:Hamilton},\ref{eq:eff}) here is based on the following circumstance: since the relation $|\vek{P}|=MV \gg |\vek{p}|=\mu v$ is satisfied in the problem under consideration, we can consider the motion of a heavy atom as a whole as the motion of a classical particle. At the same time, the dynamics of a light electron relative to a proton is described by a quantum equation. An additional justification for the applicability of this approach is the well-known fact that the gas laws are perfectly described down to fairly low temperatures within the classical model of Maxwell-Boltzmann ideal gas. In a set of works \cite{MelSchm,Melezhik2001,MelezhikCohen, Melezhik2009,Melezhik2021}, the quantum-quasiclassical approach was successfully applied to quantitatively describe various processes in different areas. The key idea of this approach goes back to the works \cite{Flannery1,Flannery2,Billing}, where it was applied to the molecular dynamics.

To integrate equations (\ref{eq:Schr}),(\ref{eq:Hamilton}) simultaneously, we need to set the initial conditions at $t=0$ defined by the physics of the problem. We have considered the case of the dynamics of a hydrogen atom under the action of a laser pulse (1), resting ($\vek{P}_0=0$) before the interaction is turned on at $t=0$ at the origin ($\vek{R}_0=0$) in the ground state $\phi_{100}(\vek{r})$
\begin{align}
\label{eq:atom0}
\psi(\vek{r},t=0)= \phi_{100}(\vek{r})\,,
\end{align}
\begin{align}
\label{eq:CM}
\vek{R}(t=0)= \vek{R}_0\,,\,\,\,\,\vek{P}(t=0)=\vek{P}_0\,.
\end{align}

We integrate the time-dependent Schr\"odinger equation (\ref{eq:Schr}) by applying the 2D discrete-variable representation (DVR)~\cite{dvr3,dvr1,SM2023} simultaneously with the Hamilton equations (\ref{eq:Hamilton}) by the St\"ormer-Verlet method~\cite{Verlet} adapted in~\cite{Melezhik2021,Mel2023} for the quantum-quasiclassical case (see Appendix B).

As a result of the integration of the hybrid system of equations (\ref{eq:Schr}),(\ref{eq:Hamilton}), the wave packet $\psi(\vek{r},t)$ and the atom CM trajectory $\vek{R}(t)$ with its momentum $\vek{P}(t)$ are calculated in the time interval $0\leq t\leq T_{max}$, where the end of integration $T_{max}$ can exceed the time of pulse duration $T_{out}$. Then, we can also calculate the excitation and ionization probabilities by the laser pulse~\cite{MS2023}, and analyse the acceleration and twisting of the atom.

\section{RESULTS and DISCUSSION}
\subsection{Excitation and ionization}

\begin{figure*}
\centering\includegraphics[scale=0.6]{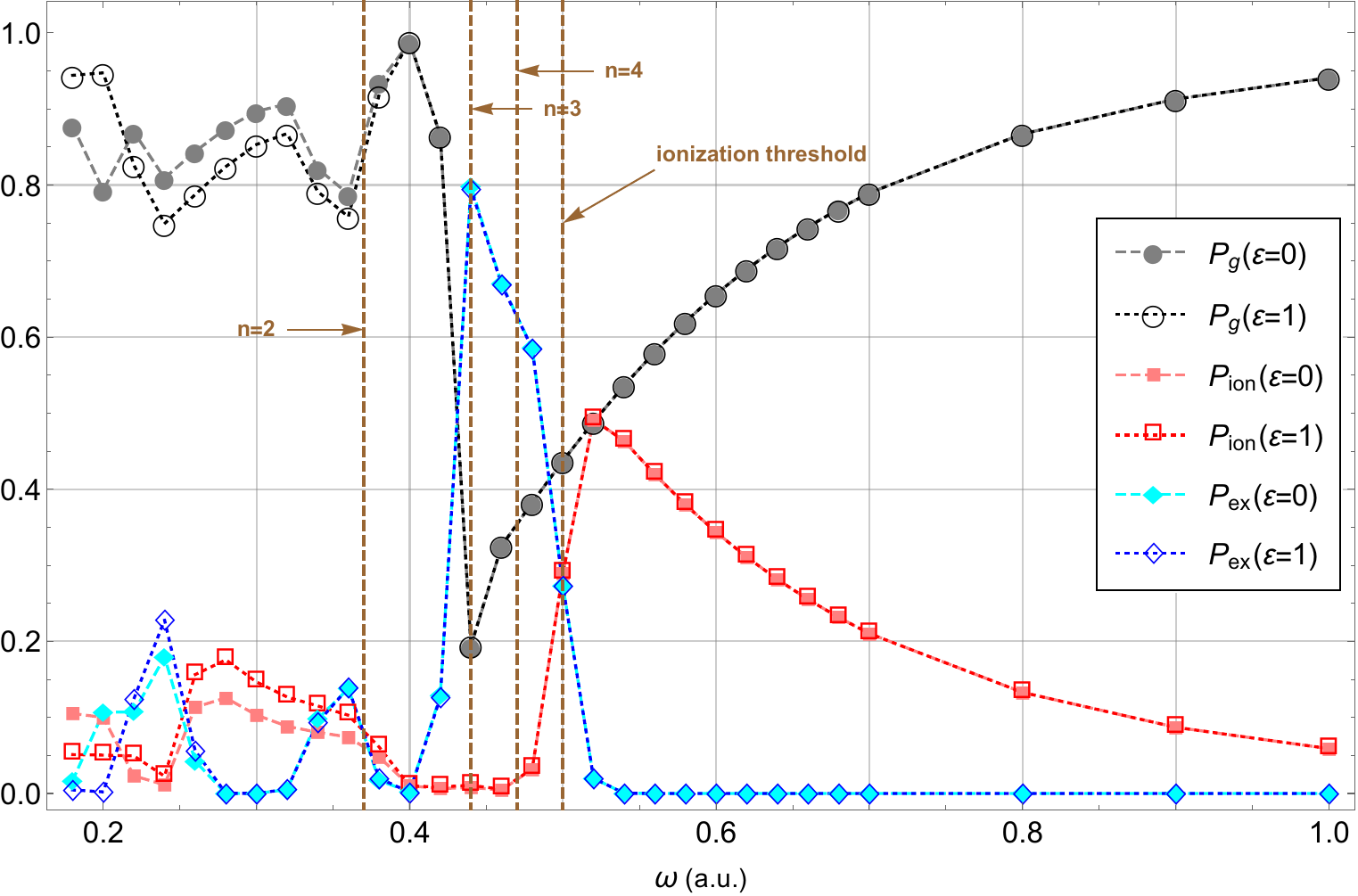}
\caption{(color online) The calculated dependencies on $\omega$ of the ground-state probability $P_g(\omega)$ and the probabilities
of atomic excitation $P_{ex}(\omega)$ and ionization $P_{ion}(\omega)$ for the laser intensity $10^{14}$ W/cm$^2$ and 7.6 fs
pulse duration for linear (fully colored symbols) and circular (open symbols) laser polarizations.}
\label{fig:1}
\end{figure*}

In our previous work~\cite{MS2023}, we investigated the acceleration of the hydrogen atom by linearly polarized laser pulses of intensity I=10$^{14}$ W/cm$^2$ and duration $\sim$ 8fs ($0\leq t
\leq T_{out}=NT=100\pi$a.u.$\approx$7.6fs) in the frequency range 0.15a.u.$\lesssim\omega\lesssim$ 1a.u. (5ev$\lesssim\hbar\omega\lesssim$27eV), as well as the excitation and ionization of the atom in this range. We have found a strong correlation between the total probability of excitation and ionization of an atom and its acceleration (the magnitude of the atom momentum $|\vek{P}(t=T_{out})|=MV_z$ after the action of the laser pulse). Here we extend the investigation to elliptically polarized laser fields. In Figure \ref{fig:1}, we present the calculated dependencies of the excitation $P_{ex}(\omega)$ and ionization $P_{ion}(\omega)$ probabilities for the atom in the linearly and circularly polarised fields as well as the populations $P_g(\omega)$ of the ground state of the atom at the end of the laser pulse $t=T_{out}=NT=100\pi$a.u.. Note that the pulse duration $T_{out}$ was fixed for varying laser frequency $\omega=2\pi/T$ by varying the number of optical cycles $N$ as in ~\cite{MS2023}. The values $P_{g}(\omega)$ were obtained by projecting the calculated electron wave-packet $\psi(\vek{r},\omega,t=T_{out})$ at the end of the laser pulse on the unperturbed ground state $\phi_{100}(\vek{r})$ of the hydrogen atom
\begin{align}
\label{eq:ex1}
P_g(\omega)=\mid \langle\psi\mid\phi_{100}\rangle\mid^2=\mid\int\psi(\vek{r},\omega,T_{out})\phi_{100}(\vek{r})d\vek{r}\mid^2\,.
\end{align}
To calculate the excitation probability $P_{ex}(\omega)=\sum_{n=2}^{\infty}P_n(\omega)$ of an atom by a laser pulse , we used the procedure suggested in our previous work~\cite{SM2023}. The idea consists in the following. The populations $P_n(\omega)$ of $2\leq n\leq 9$ states were calculated in the same way as the population of the ground state (\ref{eq:ex1}). To evaluate the remaining infinite sum $\sum_{n=9}^{\infty}P_n(\omega)$ in the total excitation probability $P_{ex}(\omega)$, we applied the ``interpolation'' procedure proposed in ~\cite{SM2023}. The ionization probability of the atom by the laser pulse $P_{ion}(\omega)$ was calculated by the formula $P_{ion}(\omega)=1-P_g(\omega)-P_{ex}(\omega)$.

\begin{figure*}
\centering\includegraphics[scale=0.6]{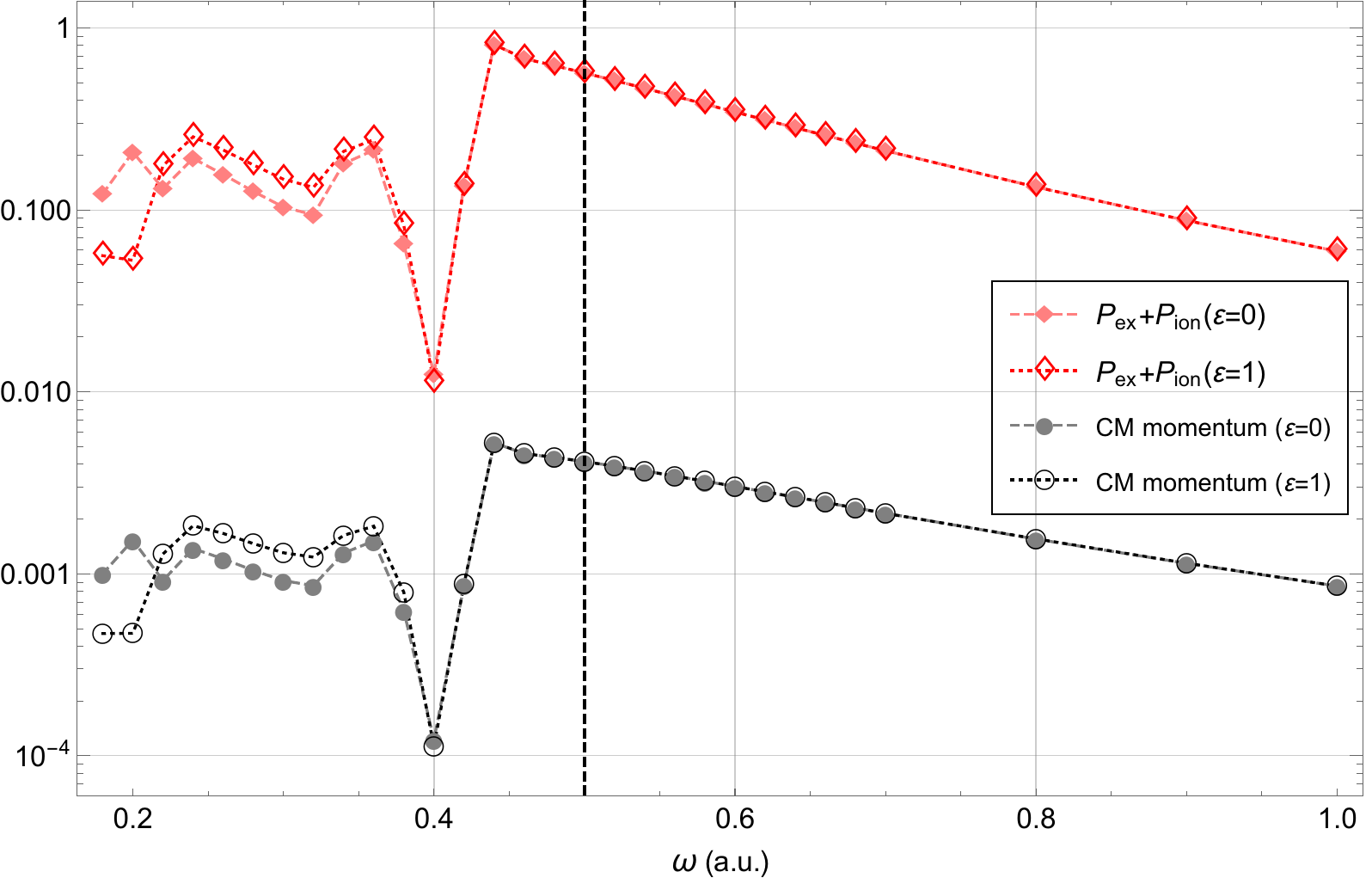}
\caption{(color online) The calculated dependencies on $\omega$ of the total excitation and ionization probability $P_{ex}(\omega,\varepsilon)+P_{ion}(\omega,\varepsilon)$, and the momentum $P_z(\omega,\varepsilon)=MV_z(\omega,\varepsilon)$ of the CM of the atom for the laser intensity $10^{14}$ W/cm$^2$ and 7.6 fs
pulse duration for linear (fully colored symbols) and circular (open symbols) laser polarizations. The CM momentum is given in a.u..}
\label{fig:2}
\end{figure*}

The calculation shows that changing the polarization of the laser pulse from linear ($\varepsilon=0$) to circular ($\varepsilon=1$) polarization noticeably affects the ionization and excitation of the atom only in the frequency range $\omega \lesssim $ 0.375a.u. below the excitation threshold of the state $n=2$. Moreover, in this region the laser polarization has the strongest effect on the atom ionization $P_{ion}(\omega)$. The transition from linear to circular polarization considerably increases the probability of ionization in the frequency range 0.22a.u.$\lesssim\omega\lesssim$0.375a.u., with the exception of $\omega=$0.24a.u., the point of resonant two-photon transition from $n=1$ to $n'=4$~\cite{MS2023}, where $P_{ion}(\omega=0.24)$ practically does not depend on the pulse polarization. However, at this point the probability of excitation of the atom $P_{ex}(\omega=0.24)$ due to the transition to circular polarization increases noticeably. It should also be mentioned that with circular polarization of the pulse the processes of excitation and ionization of the atom are essentially suppressed at $\omega\lesssim$0.2a.u..

\subsection{Acceleration of neutral atoms}

Figure \ref{fig:2} shows the calculated dependence on the laser frequency and its polarization of the momentum $\vek{P}(\omega,\varepsilon,t=T_{out})=P_z(\omega,\varepsilon)\hat{{\bf z}}=M V_z(\omega,\varepsilon)\hat{{\bf z}}$ of the atom CM reached at the laser pulse end  ($t=T_{out}$) as a result of the acceleration of the atom due to its interaction with the laser field. At the same time, the components $V_x$ and $V_y$ of the CM velocity in the laser polarization plane are negligibly small compared to the CM velocity in the direction of propagation of the laser pulse $V_z$, achieved at the end of its action.
The calculated curves of the dependence of the total probability of excitation and ionization of the atom $P_{ex}(\omega,\varepsilon)+P_{ion}(\omega,\varepsilon)$ on the laser frequency and its polarization are also given here, which in detail repeat the shape of the curves $P_z(\omega,\varepsilon)$ ($V_z(\omega,\varepsilon)$). That is, regardless of polarization, we observe a strict correlation between the total probability $P_{ex}(\omega,\varepsilon)+P_{ion}(\omega,\varepsilon)$ of excitation and ionization of the atom and the achieved velocity $V_z(\omega,\varepsilon)$ of the atom in the z-direction of propagation of the laser pulse as a result of the acceleration of the atom during its interaction with the alternating laser field.

The calculation carried out confirms the mechanism of acceleration of the atom CM established in our work~\cite{MS2023} for a linearly polarized laser pulse. The generation of a nonzero dipole moment between the proton and the electron cloud, which under the action of a laser pulse passes either to the excited state of the atom or to its continuum, is the cause of the acceleration of the CM of the atom. The calculation shows that, as in the case of excitation and ionization, the influence of laser polarization on the acceleration of the atom is noticeable only in the frequency range $\omega \lesssim $0.375 a.u.. Moreover, at frequencies 0.22a.u.$\lesssim\omega \lesssim$ 0.375 a.u. the transition to circular polarization increases the acceleration of the atom and at $\omega \lesssim$ 0.22a.u. leads to its decrease in comparison with the results for linear polarization.

Since the laser polarization does not have a significant effect on the acceleration of the atom and its probability of excitation and ionization, the frequency regions of two-photon ($n=1 \rightarrow n'=3,4$) $\omega\sim $(0.22-0.24)a.u. and one-photon ($n=1 \rightarrow n'=3-5$) $\omega\sim $(0.42-0.48)a.u. resonance transitions established in the work~\cite{MS2023} for linear polarization as the most promising for accelerating atoms also retain their prospective for circular polarization.

\begin{figure}
\parbox{0.5\textwidth}{
\centering\includegraphics[scale=0.5]{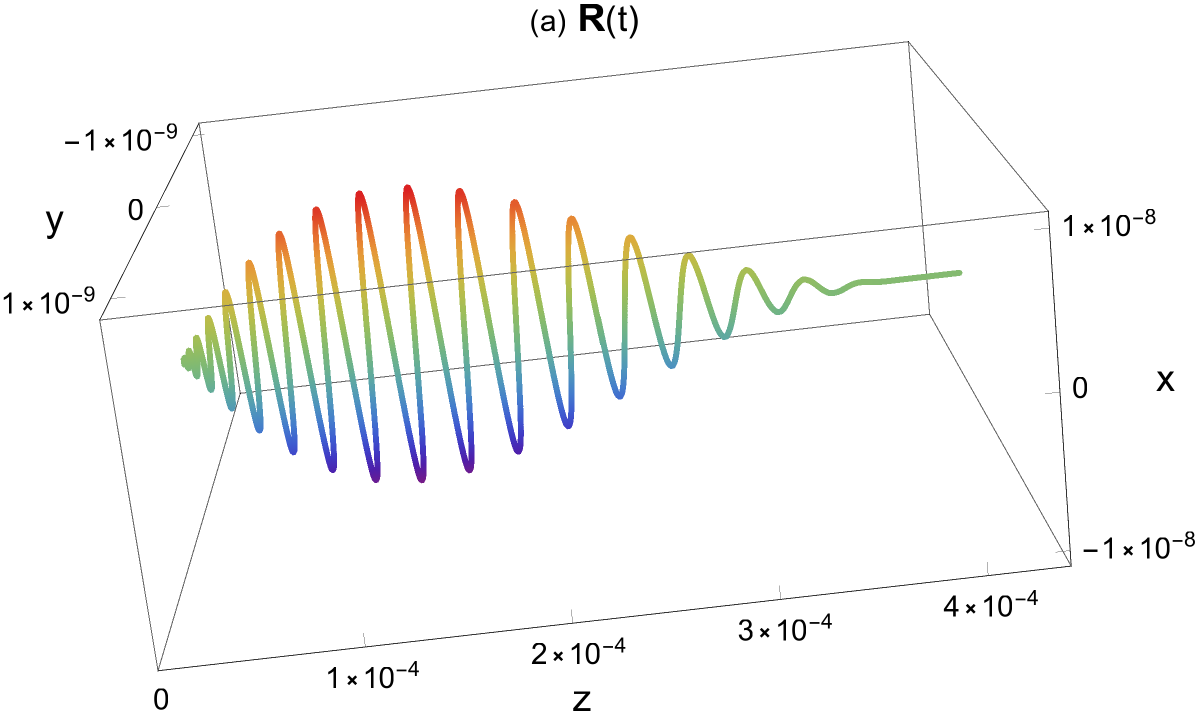}
\vspace{0.4cm}}
\parbox{0.5\textwidth}{
\centering\includegraphics[scale=0.5]{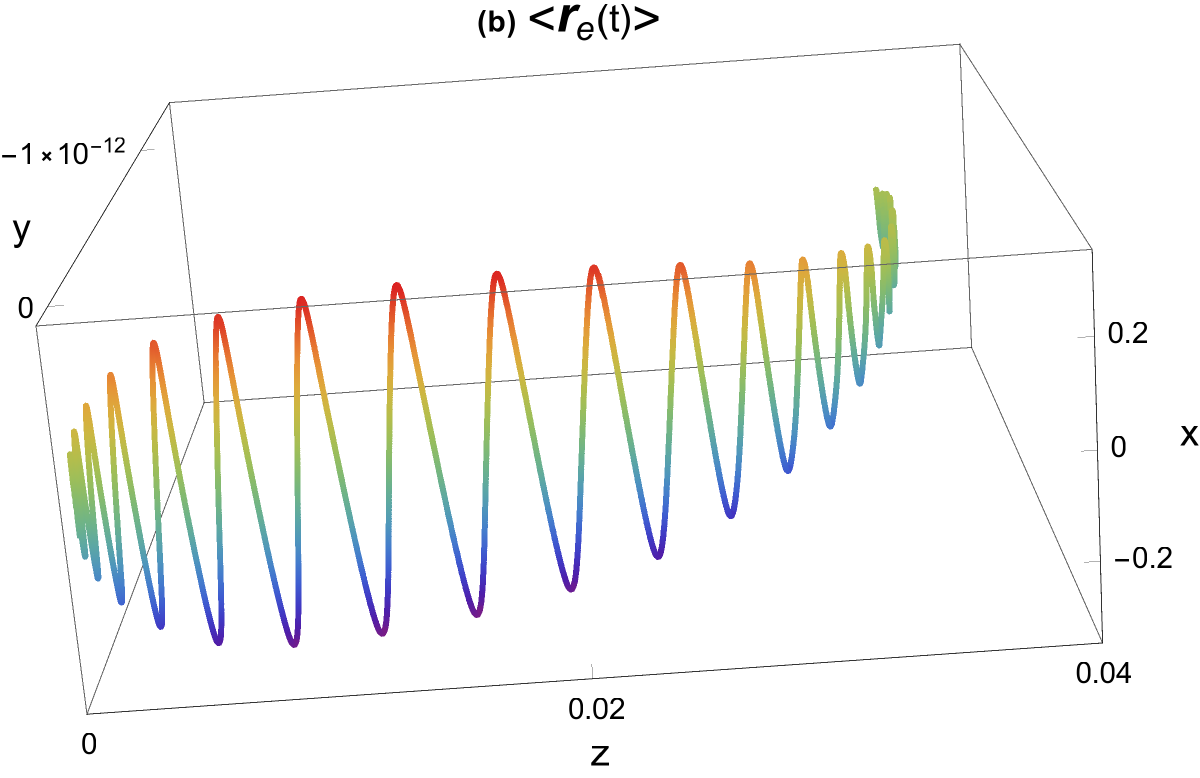}
\vspace{0.4cm}}
\parbox{0.5\textwidth}{
\centering\includegraphics[scale=0.5]{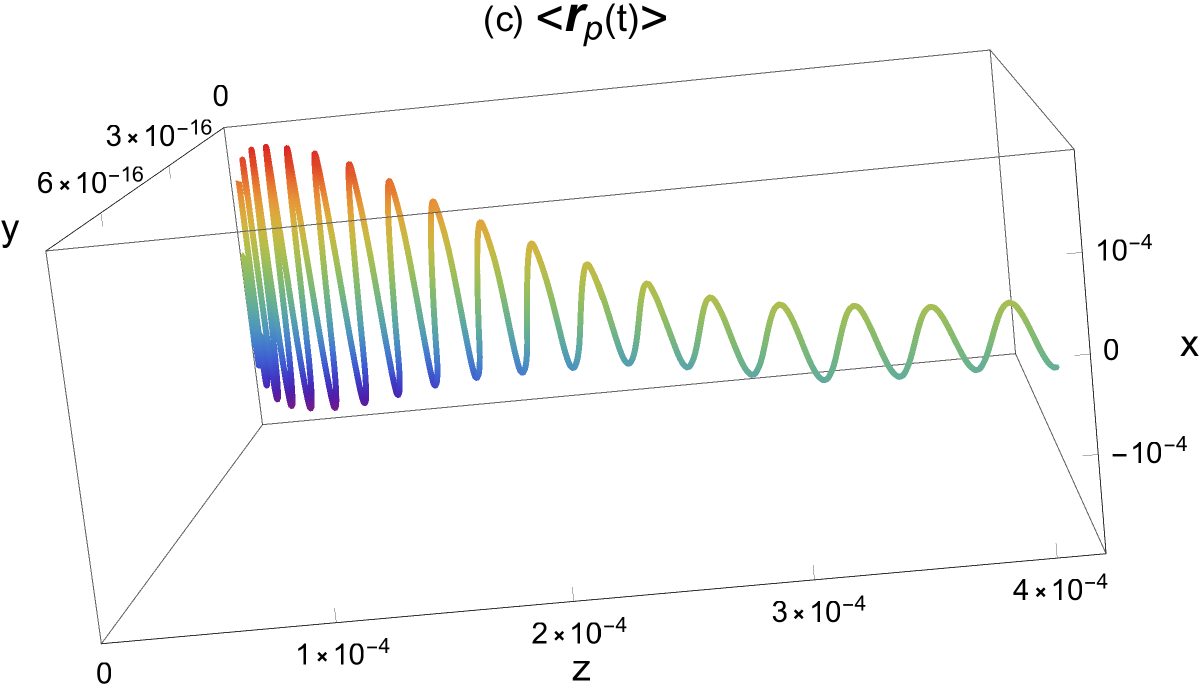}
\vspace{0.4cm}}
\caption{(color online) The calculated trajectories of the hydrogen atom center-of-mass $\vek{R}(t)$ (a), the electron cloud $\langle\vek{r}_e(t)\rangle$ (b), and the proton $\langle\vek{r}_p(t)\rangle$ (c) in the process of interaction with a linearly polarized laser pulse ($\varepsilon=0$) of 10$^{14}$ W/cm$^2$ intensity with $\omega=0.48$ a.u. and   $T_{out}=$7.6fs duration. The calculation have been performed on the time interval $0 \leq t\leq T_{max}=104\pi$a.u.=7.9fs.}
\label{fig:3}
\end{figure}

\subsection{Twisting of neutral atom by elliptically polarized laser field}

Our approach also allows one to investigate the possibility of twisting an accelerated atom when it interacts with a laser pulse. To investigate this possibility, we have analysed the trajectories in space with time evolution of the CM of the atom $\vek{R}(t)$ as well as the proton position
\begin{align}
\langle\vek{r}_p(t)\rangle =\vek{R}(t)-\frac{m_e}{M}\langle\vek{r}(t)\rangle
\end{align}
and the ``electron position''
\begin{align}
\langle\vek{r}_e(t)\rangle =\vek{R}(t)+\frac{m_p}{M}\langle\vek{r}(t)\rangle\,,
\end{align}
where the mean value of the relative variable of the electron $\langle\vek{r}(t)\rangle$ in the left parts of the above equations is calculated by averaging over the instantaneous electron distribution in space $|\psi(\vek{r},t)|^2$ obtained by integration of  equations (\ref{eq:Schr}),(\ref{eq:Hamilton})
\begin{align}
\langle\vek{r}(t)\rangle = \langle\psi(\vek{r},t)|\vek{r}|\psi(\vek{r},t)\rangle =\int |\psi(\vek{r},t)|^2 \vek{r}d\vek{r}\,.
\end{align}

Figure \ref{fig:3} shows the results of calculating the trajectories of the atom center-of-mass $\vek{R}(t)$, the electron cloud $\langle \vek{r}_e(t)\rangle$, and the proton $\langle \vek{r}_p(t)\rangle$ in the time interval $0\leq t\leq T_{max}$=104$\pi$a.u. including the time of interaction of the atom with a linearly polarized pulse of 10$^{14}$W/cm$^2$ with $\omega$=0.48a.u.. The calculation is performed for a time-interval slightly exceeding the action time of the laser pulse $0\leq t\leq T_{max} > T_{out}$=100$\pi$a.u.. By the time the pulse ends $t=T_{out}$, the atom CM starting from the initial point $\vek{R}(t=0)$=0 reaches the point $Z$=3.9$\times$10$^{-4}$ a.u.. Further, after the laser pulse has finished, the CM of the atom moves along the $z$-axis with a constant speed $V_z\simeq$ 5.2 m/sec. A small deviation observed in the transverse directions of the variables $\langle \vek{r}_p(t)\rangle$ and $\langle \vek{r}_e(t)\rangle$ with the laser pulse attenuation is explained by the spreading of part of the electron  wave packet that appeared in the continuum as a result of the atomic ionization. However, the mentioned effect is practically negligible here because the laser frequency $\omega$=0.48a.u. is chosen from the frequency region where the probability of ionization of the atom (equal to the probability for atom transition to continuum) is strongly suppressed to $P_{ion}(\omega=0.48)\simeq3.2\times$10$^{-2}$ (see Fig. \ref{fig:1}). We see that with linear polarization ($\varepsilon$=0), the twisting of the atom in the plane orthogonal to the direction of the atom acceleration does not occur.

\begin{figure}
\parbox{0.5\textwidth}{
\centering\includegraphics[scale=0.5]{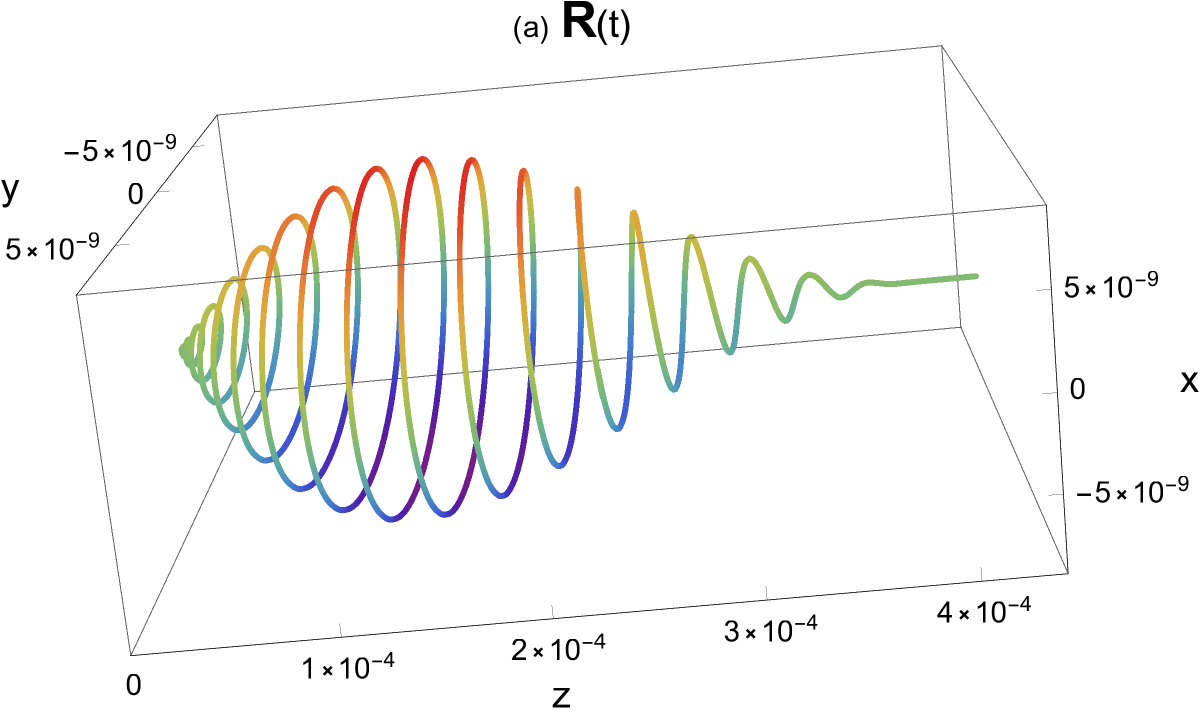}
\vspace{0.4cm}}
\parbox{0.5\textwidth}{
\centering\includegraphics[scale=0.5]{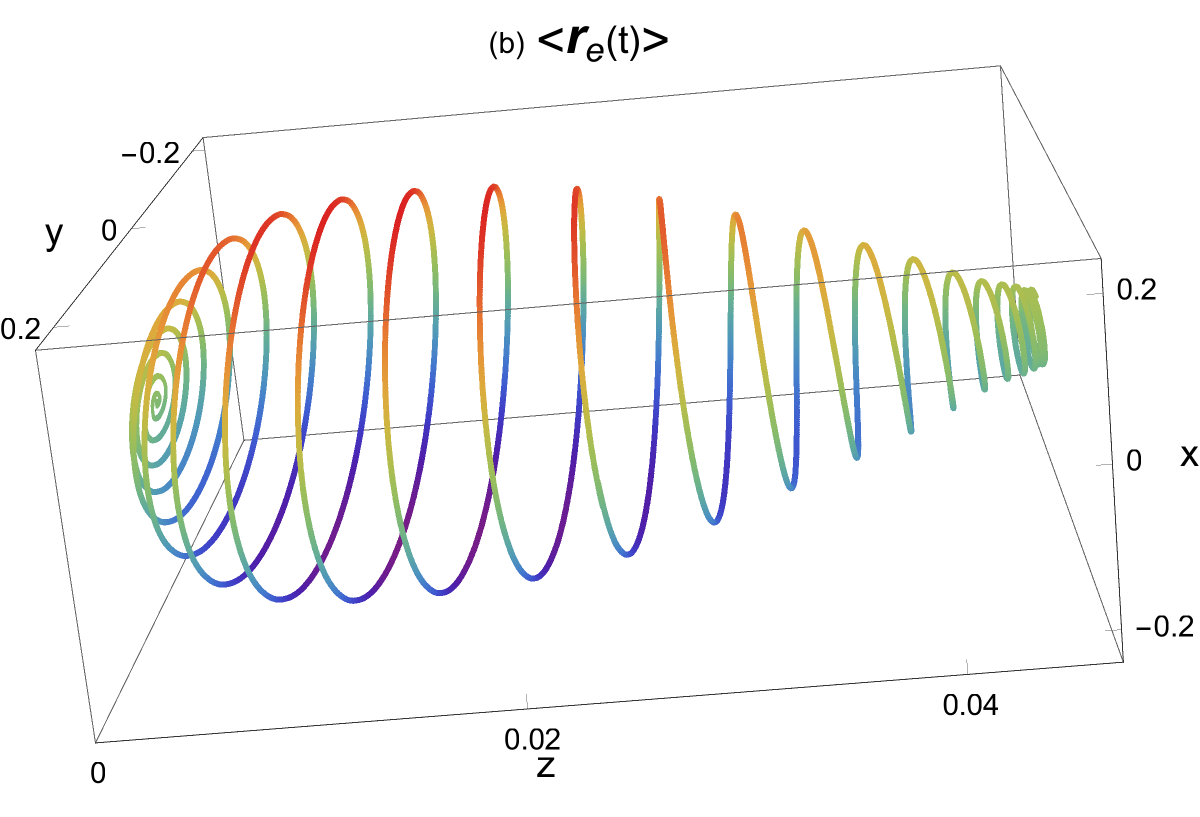}
\vspace{0.4cm}}
\parbox{0.5\textwidth}{
\centering\includegraphics[scale=0.5]{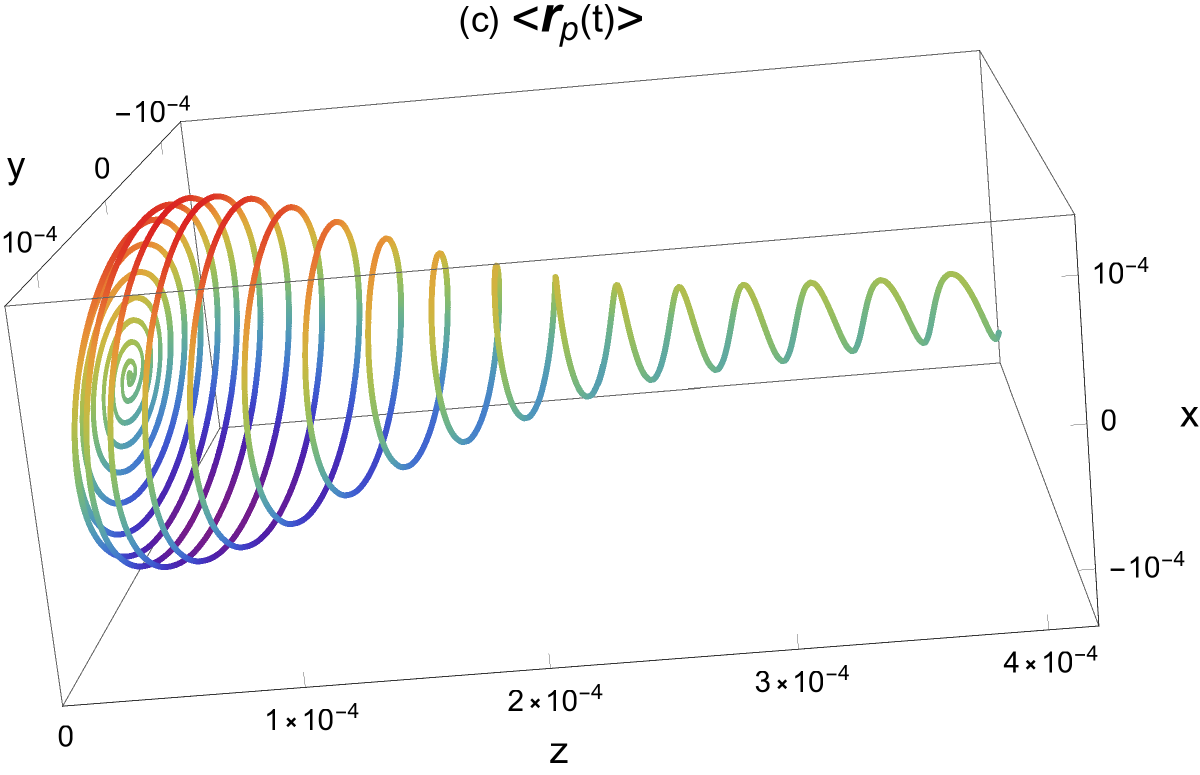}
\vspace{0.4cm}}
\caption{The calculated trajectories of the hydrogen atom center-of-mass $\vek{R}(t)$ (a), the electron cloud $\langle\vek{r}_e(t)\rangle$ (b), and the proton $\langle\vek{r}_e(t)\rangle$ (c) in the process of interaction with a circularly polarized laser pulse ($\varepsilon=1$) of 10$^{14}$ W/cm$^2$ intensity, with $\omega=0.48$a.u. and   $T_{out}=$7.6fs duration. The calculation has been performed in the time interval $0 \leq t\leq T_{max}=104\pi$a.u. =7.9fs.}
\label{fig:4}
\end{figure}

\begin{figure*}
\centering\includegraphics[scale=0.6]{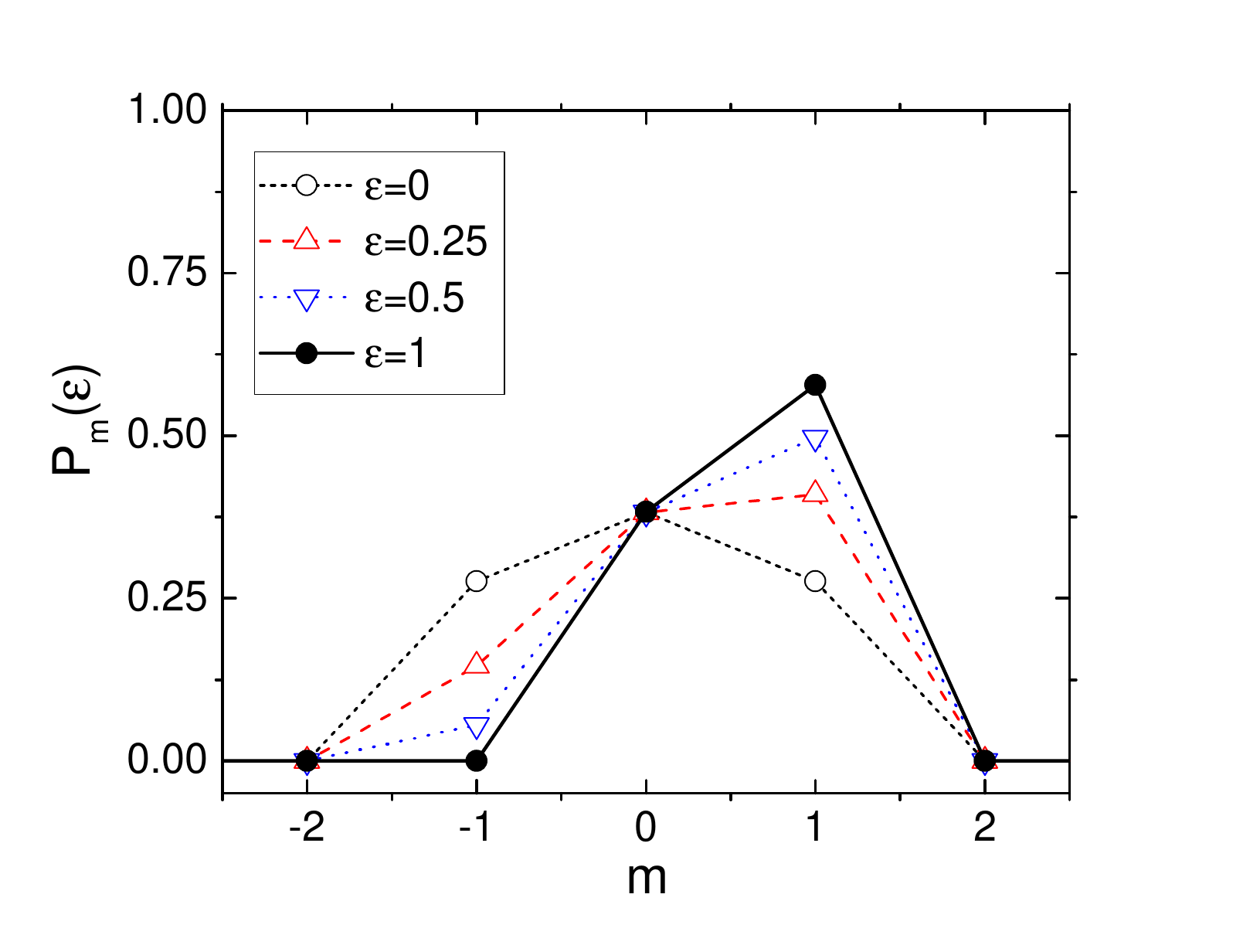}
\caption{(color online) The calculated dependencies on the ellipticity $\varepsilon$ for the populations $P(m)$ of the orbital angular momentum projection $m$ in the direction of the laser pulse propagation at the end of the pulse duration 7.6fs for the laser with intensity of 10$^{14}$ W/cm$^2$ and frequency $\omega=0.48$a.u..}
\label{fig:5}
\end{figure*}

Next, we have performed similar calculations for the elliptical polarization of laser radiation. Figure \ref{fig:4} shows the results of calculation of the trajectories of the atom center-of-mass $\vek{R}(t)$, the electron cloud $\langle \vek{r}_e(t)\rangle$ and the proton $\langle \vek{r}_p(t)\rangle$ during the interaction of the atom with a circularly polarized pulse ($\varepsilon$=1). Figure \ref{fig:4}(a) shows the calculated trajectory of the CM of the atom $\vek{R}(t)$ during the same time interval like above $0\leq t\leq T_{max}$=104$\pi$a.u., which slightly exceeds the time of atom interaction with a laser pulse $T_{out}$. Here, one can see the dynamics of the twisting of the atom CM during its interaction with the laser pulse relative to the direction of the pulse propagation (z-axis) and its gradual exit to a linear trajectory with pulse attenuation. At the same time, both the electron cloud (Fig. \ref{fig:4}(b)) and the proton (Fig. \ref{fig:4}(c)) are twisting (but in different directions - clockwise and counterclockwise) relative to the $z$-axis. Moreover, unlike the CM of the atom, the twisting of the electron cloud and the proton relative to the direction of propagation of the laser pulse is preserved for some time after its attenuation. (At the end of the pulse, the CM of the atom, the electron and the proton reach points $Z$=3.9$\times$10$^{-4}$a.u., $z_e$=0.046a.u. and $z_p$=3.7$\times$10$^{-4}$a.u., respectively.) To clarify the mechanism of twisting of the electron cloud of the hydrogen atom by an elliptically polarized laser pulse, we have calculated the populations $P_m(\varepsilon)$ of the states of the hydrogen atom with different $m$ (projection of the electron angular momentum on the direction of the laser pulse propagation which coincides with the direction of the atom CM motion after the pulse termination) at the end of the pulse for different ellipticities $\varepsilon$
\begin{align}
\label{eq:hel}
P_m(\varepsilon)= \sum_{n=l+1}^{n_{max}}\sum_{l=\mid m\mid}^{n_{max}-1}\mid\langle\psi(T_{max},\varepsilon)\mid\phi_{nlm}\rangle\mid^2
=\sum_{n=l+1}^{n_{max}}\sum_{l=\mid m\mid}^{n_{max}-1}\mid\int\psi^*(\vek{r},T_{max};\varepsilon)\phi_{nlm}(\vek{r})d\vek{r}\mid^2\,,
\end{align}
where the choice of the summation limit $n_{max}$=10 was determined by the number of completely filled $l$-shells taken into account in the DVR approximation~\cite{SM2023} of the angular part of the electron wave packet $\psi(\vek{r},T_{max},\varepsilon)$ calculated at $t=T_{max}$. In formula (\ref{eq:hel}) the eigenfunctions $\phi_{nlm}(\vek{r})$ of the discrete spectrum of the hydrogen atom are used, in which the $z$-axis is chosen as the quantization axis coinciding with the direction of propagation of the laser pulse.
In Figure \ref{fig:5} we present the result of the calculations of the populations of the atomic states with different $m$ performed for different $\varepsilon$. In the case of linear polarization ($\varepsilon=0$), the states of the atom with the projections $m=1$ and $-1$ of its orbital angular momentum onto the direction of its motion after the termination of the laser pulse are populated with the same nonzero probability $P_1(0)=P_{-1}(0)=0.27$. However, the total value of the projection of the orbital angular momentum $\hat{\vek{l}}$ on the direction of the pulse propagation
\begin{align}
\langle \hat{l}_z \rangle=\sum_m P_m(\varepsilon) m
\end{align}
 is equal 0 in this case.
This apparent contradiction is removed by the well-known fact about the possibility of representing a linearly polarized wave as the sum of two circularly polarized waves with left and right polarization. The calculation performed demonstrates that for elliptical polarization ($0<\varepsilon\leq 1$) with increasing $\varepsilon$ the population of states $P_m(\varepsilon)$ with positive $m$ increases in contrast to the population of states with negative $m$. The effect reaches its maximum value $P_1(\varepsilon)=0.55$ for circular polarization ($\varepsilon=1$). In this case, the projection of the electron angular momentum on the direction of the atom motion is positive and reaches the value $\langle \hat{l}_z\rangle=0.55$. It is clear that to obtain a negative projection of the electron angular momentum on the momentum $\vek{P}$ of atom center-of-mass, it is necessary to change the direction of rotation of the vectors $\vek{E}$ and $\vek{B}$ in the $xy$ plane to the opposite one (i.e., to change the sign of the ellipticity $\varepsilon$ to the negative one). It should also be noted that the population of the state with $m=0$ does not depend on the ellipticity value but is determined by other parameters of the laser pulse: the radiation intensity $I$, its frequency $\omega$, the shape of the laser pulse $f(t)$ and its duration. The ellipticity $\varepsilon$ determines the relative populations of the states of the atom with different nonzero $m$. The necessary condition for creation of the twisted atoms, i.e. for the creation of the atoms with a nonzero projection of the electron angular momentum in the direction of motion of the atom, is the demand for $\varepsilon\neq 0$.


\section{CONCLUSION}

We have investigated theoretically the acceleration and twisting of hydrogen atoms as well as their excitation and ionization  by elliptically polarized laser pulses of 10$^{14}$ W/cm$^2$ in the frequency region 0.15 a.u.$\leq \omega \leq$1 a.u. (5 eV $\leq\hbar\omega \leq$ 27 eV) near the ionization threshold ($\omega=$0.5a.u.) with $\simeq$ 8fm duration. The study was performed within the framework of a hybrid quantum-quasiclassical approach in which the time-dependent Schr\"odinger equation for an electron
and the classical Hamilton equations for the CM of an atom are simultaneously integrated.

We have found a strict correlation between the total probability of excitation and ionization of an atom and the achieved velocity of the atom in the direction of propagation of the laser pulse as a result of the acceleration of the atom during its interaction with the elliptically polarized laser pulse. This effect was initially observed in our previous work~\cite{MS2023} for linearly polarized fields. Here we have shown that the ellipticity noticeably affects the excitation and ionization probabilities as well as the atom acceleration only in the limited range of laser frequencies 0.22a.u $\lesssim\omega\lesssim$0.375 a.u.. However, the found strong correlation between the atom velocity achieved at the pulse termination and the total excitation and ionization probabilities remains for any ellipticity.

We have also shown that the elliptically polarized laser pulse together with the acceleration can also twist the atom. The twisting occurs with the appearance of laser ellipticity, increases with ellipticity and  reaches a maximal value for circular polarization of a laser pulse.

Based on the performed investigation, we propose that using elliptically polarised short-wavelength short-range laser pulses have good prospects for obtaining accelerated and twisted atoms, the production of which is of interest for a number of promising applications. Thus, among the applications under discussion, one can note projects on using accelerated atoms for lithography of microchips in microelectronics and for plasma diagnostics in TOKAMAKS~\cite{Cai}. It is also proposed to use twisted atoms as a new probe for investigations using an additional degree of freedom of orbital angular momentum~\cite{Luski}.

\acknowledgements

The authors thank D.V. Karlovets, V.I. Korobov, Yu.V. Popov, and O.V. Teryaev for fruitful discussions.  The work was supported by the Russian Science Foundation under Grants No. 20-11-20257.

\appendix

\section{Non-dipole interaction of hydrogen atom with laser field}

When deriving formulas (\ref{eq:v1}),(\ref{eq:v2}) for the interaction potential (\ref{eq:v0}) of a hydrogen atom with a laser field taking into account non-dipole corrections of the order of $1/c$, we used the expansion of the electric and magnetic fields in this small parameter with an accuracy of up to terms  $\sim 1/c$ inclusive
\begin{eqnarray}
{\bf E} &=& E_0(t)\Bigg\{\Big[({\bf\hat{x}}\cos(\omega t)+\varepsilon{\bf\hat{y}}\sin(\omega t))
+\frac{\bar{f}(t)}{2Nf(t)}
({\bf\hat{x}}\sin(\omega t)-\varepsilon{\bf\hat{y}}\cos(\omega t))\Big]  \nonumber\\
&+& \frac{\omega}{c}z\Big[
({\bf\hat{x}}\sin(\omega t)-\varepsilon{\bf\hat{y}}\cos(\omega t))
-\frac{\bar{f}(t)}{2Nf(t)}
({\bf\hat{x}}\cos(\omega t)+\varepsilon{\bf\hat{y}}\sin(\omega t))\Big]
\Bigg\}\,,
\end{eqnarray}
\begin{eqnarray}
{\bf B}=\frac{1}{c}E_0(t)\Big[{\bf\hat{y}}\cos(\omega t)-\varepsilon{\bf\hat{x}}\sin(\omega t)\Big]\,.
\end{eqnarray}
Substituting the above expressions for ${\bf E}$ and ${\bf B}$ into the formula for the Lorentz force acting in the external electromagnetic field of the laser on the electron and proton in the hydrogen atom and then using the well-known relation $\vek{F}(\vek{r})=-\nabla V(\vek{r})$ connecting the vector field with a scalar potential field, we obtain the potentials $V_e(\vek{r}_e)$ and $V_p(\vek{r}_p)$ describing the interaction of the electron and proton with the laser field in the non-dipole approximation.

Then representing the interaction potential of the hydrogen atom with the laser field as the sum $V_e(\vek{r}_e) +V_p(\vek{r}_p)$ and passing to the coordinates of the center-of-mass $\vek{R}$ and the relative motion $\vek{r}$ in the hydrogen atom
\begin{align}
\vek{R}=\frac{m_e}{M}\vek{r}_e+\frac{m_p}{M}\vek{r}_p\approx \vek{r}_p\,\,\,\,\,\,\,\, \vek{P}=\vek{p}_e+\vek{p}_p\,,
\end{align}

\begin{align}
\vek{r}=\vek{r}_e-\vek{r}_p\,\,\,\,\,\,\,\, \vek{p}=\frac{m_p}{M}\vek{p}_e-\frac{m_e}{M}\vek{p}_p\approx \vek{p}_e\,\,,
\end{align}
with neglecting terms of the order of $1/M=(m_e+m_p)^{-1}$, we finally obtain the interaction potential (\ref{eq:v0}) of the hydrogen atom with the elliptically polarized laser field (1) defined in the non-dipole approximation with an accuracy of the order of $1/c^{2}$ and $1/M$ by formulas (\ref{eq:v1}),(\ref{eq:v2}).

\section{Quantum-quasiclassical computational scheme}

To integrate the coupled system of equations (\ref{eq:Schr}),(\ref{eq:Hamilton}), a special computational scheme was applied.
Here for the numerical integration of the 3D equation (\ref{eq:Schr}) we use the computational scheme developed in our work~\cite{SM2023} for the 3D time-dependent Schr\"odinger equation  describing the hydrogen atom in strong elliptically polarized fields. It is based on a 2D DVR~\cite{dvr3,dvr1} for approximating the angular part of the calculated electron wave-packet $\psi(\vek{r},t_n)$ and a tailored splitting-up procedure for realization of the propagation in time of the wave-packet $\psi(\vek{r},t_n)\rightarrow \psi(\vek{r},t_{n+1}) $~\cite{dvr3,SM2023}. Integrations were performed with the time step $\Delta t=0.01$a.u. on the radial grid with $N_r$=2000 grid points up to the radial boundary $r_m$=500a.u. and with $N_{\Omega}=N_{\theta}\times N_{\phi} =17\times 17$ Gaussian angular grid points of 2D DVR~\cite{SM2023}.

Simultaneously with the forward in time propagation
$t_n\rightarrow t_{n+1}=t_n+\Delta t$ of the electron wave-packet $\psi(\vek{r},t_{n})\rightarrow \psi(\vek{r},t_{n+1})$ when integrating the time-dependent Schr\"odinger equation (\ref{eq:Schr}), we integrate
the Hamilton equations of motion ~(\ref{eq:Hamilton}) with the second-order St\"ormer-Verlet method~\cite{Verlet} adapted to our problem~\cite{Mel2023}
\begin{equation}
\vek{P}(t_n+\frac{\Delta t}{2}) = \vek{P}(t_n) - \frac{\Delta t}{2}\frac{\partial}{\partial \vek{R}}H_{eff}(\vek{P}(t_n+\frac{\Delta t}{2}),\vek{R}(t_n))\,,\nonumber
\end{equation}
\begin{align}
\vek{R}(t_{n+1}) = \vek{R}(t_n) + \frac{\Delta t}{2}\Bigg\{\frac{\partial}{\partial \vek{P}}H_{eff}(\vek{P}(t_n+\frac{\Delta t}{2}),\vek{R}(t_n))
+\frac{\partial}{\partial \vek{P}}H_{eff}(\vek{P}(t_n+\frac{\Delta t}{2}),\vek{R}(t_{n+1}))\Bigg\}\,,\nonumber
\end{align}
\begin{equation}
\vek{P}(t_{n+1}) = \vek{P}(t_n+\frac{\Delta t}{2}) - \frac{\Delta t}{2}\frac{\partial}{\partial \vek{R}}H_{eff}(\vek{P}(t_n+\frac{\Delta t}{2}),\vek{R}(t_{n+1}))\,.
\end{equation}

In our case, when the effective classical Hamiltonian $H_{eff}(\vek{P},\vek{R})$ is defined by Eq.(\ref{eq:eff}), the formulas for implementing the St\"ormer-Verlet method take the form
\begin{eqnarray}
{P_x}(t_n+\frac{\Delta t}{2})&=&{P_x}(t_n)-\frac{\Delta t}{2}E_0(t_n)\frac{\omega}{c}\Bigg\{\left[\sin(\omega t_n)-\frac{\bar{f}(t_n)}{2Nf(t_n)}\cos(\omega t_n)\right]\langle z\rangle
-\frac{1}{\omega}\cos(\omega t_n)\langle p_z\rangle\Bigg\}  \nn \\
&=&{P_x}(t_n)+q(t_{n}) ~, \nn \\
{P_x}(t_{n+1})&=&{P_x}(t_n+\frac{\Delta t}{2})+q(t_{n+1})~, \nn \\
{X}(t_{n+1})&=&X(t_n)+\Delta t\frac{P_x(t_n+\frac{\Delta t}{2})}{M}\,,
\end{eqnarray}

---------------------------
\begin{eqnarray}
{P_y}(t_n+\frac{\Delta t}{2})
&=&{P_y}(t_n)+\frac{\Delta t}{2}E_0(t)\frac{\varepsilon\omega}{c}\Bigg\{\left[\cos(\omega t_n)+\frac{\bar{f}(t_n)}{2Nf(t_n)}\sin(\omega t_n)\right]\langle z\rangle
-\frac{1}{\omega}\sin(\omega t_n)\langle p_z \rangle\Bigg\} \nn\\
&=&{P_y}(t_n)+g(t_{n}) ~, \nn \\
{P_y}(t_{n+1})&=&{P_y}(t_n+\frac{\Delta t}{2})+g(t_{n+1})~, \nn \\
Y(t_{n+1})&=&Y(t_n)+\Delta t\frac{P_y(t_n+\frac{\Delta t}{2})}{M}\,,
\end{eqnarray}
-------------------------
\begin{eqnarray}
{P_z}(t_n+\frac{\Delta t}{2})
&=&{P_z}(t_n)-\frac{\Delta t}{2}E_0(t)\frac{\omega}{c}\Bigg\{\left[\sin(\omega t_n)-\frac{\bar{f}(t_n)}{2Nf(t_n)}\cos(\omega t_n)\right]\langle x\rangle
-\varepsilon\left[\cos(\omega t_n)+\frac{\bar{f}(t_n)}{2Nf(t_n)}\sin(\omega t_n)\right]\langle y\rangle \nn\\
&+&\frac{1}{\omega}\bigg[\cos(\omega t_n)\langle p_x\rangle +\varepsilon\sin(\omega t_n)\langle p_y\rangle\bigg] \Bigg\} \nn\\
&=&{P_z}(t_n)+h(t_{n}) ~, \nn \\
{P_z}(t_{n+1})&=&{P_z}(t_n+\frac{\Delta t}{2})+h(t_{n+1})~, \nn \\
Z(t_{n+1})&=&Z(t_n)+\Delta t\frac{P_z(t_n+\frac{\Delta t}{2})}{M}\,.
\end{eqnarray}
---------------------------

\end{document}